\documentclass[fleqn,10pt]{wlscirep}
\title{Plasma plumes produced by laser ablation of Al with single and double pulse schemes}

\author[1,*]{N Smijesh}
\author{Kavya H. Rao}
\author{D. Chetty}
\author{I. V. Litvinyuk}
\author{R. T. Sang}
\affil{Australian Attosecond Science Facility, Centre for Quantum Dynamics, Griffith University, Nathan, QLD-4111, Australia.}

\affil[*]{s.nadarajanachary@griffith.edu.au}


\begin{abstract}
We generated and characterized plasma with single and double picosecond laser pulses in order to study the plume dynamics and to control the plasma properties. The double-pulse scheme was found to be superior for the generation of a homogeneous plasma. The lateral expansion was prominent for irradiation schemes wherein energy of the first pulse is lower/equal to that of the second pulse. While the velocities of the fast and slow species were found to be nearly equal, the emission counts corresponding to slow species are larger for single pulse compared to the double pulse. %Also, the plume shape resembles a nanosecond laser produced plasma for DP. 
\end{abstract}
\begin{document}

\flushbottom
\maketitle
\thispagestyle{empty}

%\noindent Please note: Abbreviations should be introduced at the first mention in the main text – no abbreviations lists. Suggested structure of main text (not enforced) is provided below.

\section*{Introduction}

Laser produced plasmas (LPP) are of great importance in a variety of applications including EUV generation\cite{EUV1}, high-order harmonic generation\cite{ganeev2013high,Ganeev:12,Hutchison:12,Ganeev:07}, attosecond pulse generation\cite{Ma:16,Liu:13,PhysRevLett.93.163901}, wake-field acceleration\cite{wakefield}, nanoparticle and nanocluster generation\cite{PhysRevB.71.033406, nanoparticleapl} etc. LPPs are highly transient such that the parameters of the plume vary with time and space rapidly and therefore characterization of the plume is necessary for the above mentioned applications. The nature of expansion of LPP is essentially dependent on number density and temperature of the plume, which depends on parameters of the laser such as wavelength, pulse duration, energy and spot size, nature and pressure of the ambient gas along with material properties\cite{hassa2016}. Attaining a suitable number density at a specific high temperature is crucial for many applications, and it is challenging to achieve\cite{icf}.  Various irradiation schemes including single-pulse (SP) and double-pulse  (DP) methods \cite{dpf} were also employed previously to investigate parameters of plasmas. It has been reported that, while the temperature increases very slightly (usually $<$ 10\%), the line emission intensity from various species increases\cite{sattmann1995laser,stOnge} for the optimum delay, $\sim$500-1000 picoseconds (ps), between pulses\cite{pinon2009optical} for DP scheme.  However, it is theoretically predicted that the temperature of the plasma could be increased up to 3 times if the preformed plasma is irradiated with a delayed pulse at $\le$ 200 ps, but it has not been experimentally demonstrated to date\cite{povarnitsyn2015molecular}. There are studies on DP using laser pulses of different wavelengths where a dramatic increase in the emission intensity and plasma temperature has been observed when an infrared (IR) laser is used to irradiate the pre-formed plasma\cite{st2002enhanced,ahmed2009comparative}. Stratis \textit{et. al}. 2001\cite{Stratis:01} used the orthogonal geometry (i.e. the plasma plume formed by the first pulse interacts with the second beam passing perpendicular to the plume expansion direction) in the pre-ablation mode and found an increase in overall size and a change in the shape of the plume. Characterization of such plumes are generally carried out using techniques such as optical emission spectroscopy (OES) and time-of-flight (TOF)\cite{harilal2017high,smijesh2016spatio,smijesh2016influence} to infer the spatio-temporal nature of the plume. Standard experimental techniques capable of uncovering the early expansion of the plasma plume are time-resolved shadowgraphy\cite{shadowgraphy,Mao,Nell} and schlieren photography\cite{schlieren}. However, these techniques cannot be used to reconstruct the hydrodynamic expansion of the plume and its radiative characteristics. Therefore time-resolved plume imaging using an intensified charge coupled device (ICCD) would be a suitable choice to study the hydrodynamic expansion along with time-resolved OES for spectral information. This work compares the plume dynamics of Al plasmas generated by single and double pulse schemes.

\section*{Experimental}

\begin{figure}[ht!]
	\centering
	\includegraphics[width=0.9\linewidth]{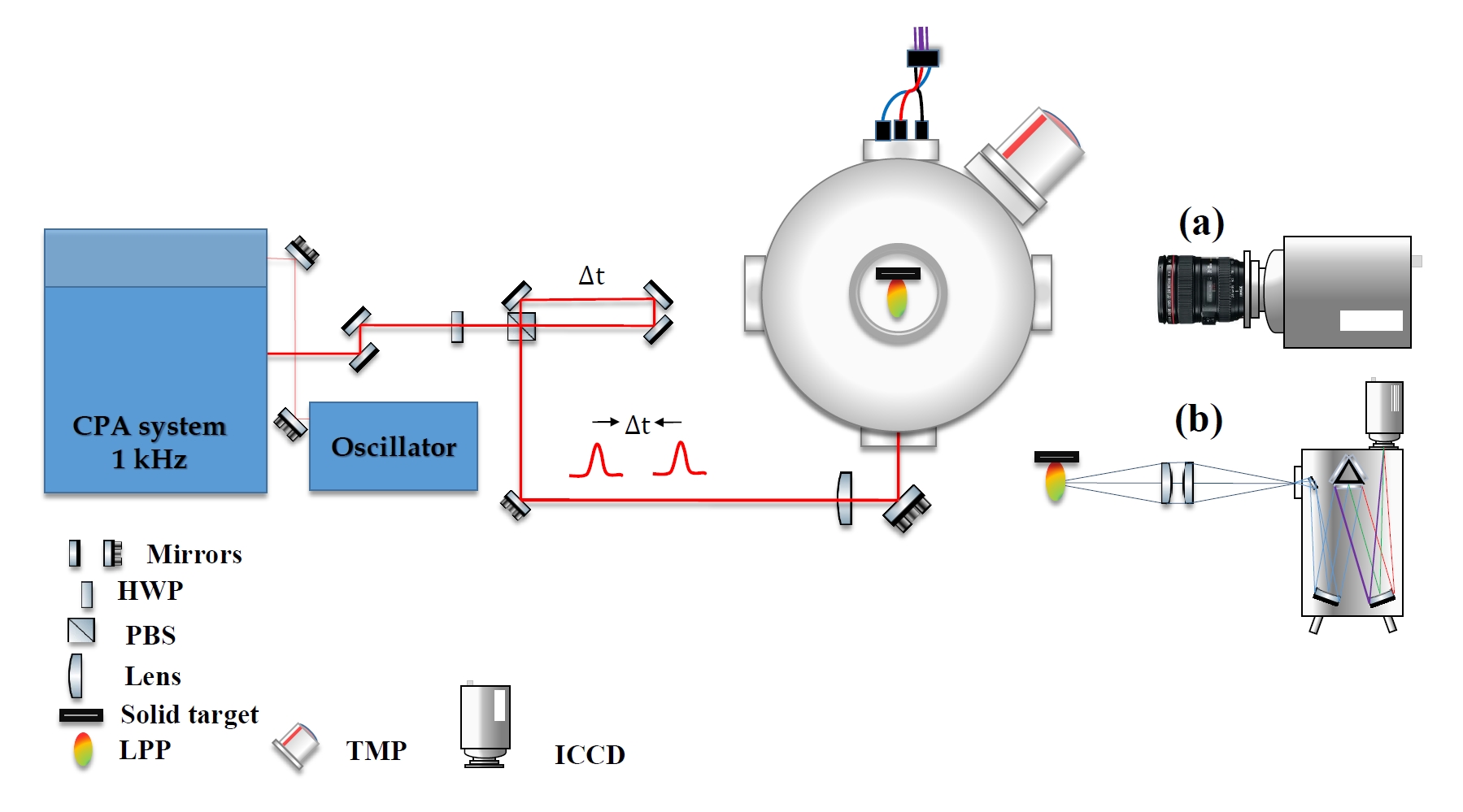}
	\caption{Experimental setup for the generation of picosecond laser plasma from solid Al target at $\sim$ 10 $^{-6}$ Torr nitrogen pressure and for (a) plasma hydrodynamics as well as (b) optical emission spectroscopic investigations.}
	\label{fig:setup}
\end{figure}

The plasma is generated using a 60 ps laser pulse at 800 nm from a mode-locked Ti: Sapphire  laser (\textit{Odin II}, Quantronix) focused to a spot size of $\sim$ 85 $\mu$m using a 500 mm plano-convex lens onto the surface of a 99.99\% pure 50 mm $\times$ 50 mm $\times$ 3 mm  Al target (\textit{ACI Alloys Inc}, USA), which is kept in nitrogen at a pressure of $\sim$ 10 $^{-6}$ Torr. The laser operates at a repetition rate of 1 kHz whereas the experiment is performed for a predefined number of irradiations on the target by positioning a fast, synchronized mechanical shutter into the beam path. The target is translated by $\sim$200 $\mu$m after each measurement to avoid repeated irradiation on the pit formed on the surface due to ablation by previous irradiations. Emissions from ions and neutrals in the range from 300 nm to 400 nm are recorded using a spectrometer (\textit{SP 2550}, Princeton Instruments) equipped with a 13 $\mu$m $\times$ 13 $\mu$m, 1024$\times$1024  Gen II ICCD (\textit{Pi: MAX 1024 f}, Princeton Instruments), with a spectral resolution of $\sim$ 0.02 nm. The layout of the experimental set up is shown in Fig. \ref{fig:setup}. Spectral lines in the emission spectra are compared with a standard NIST database \cite {NIST} and emissions from neutral as well as ionized species are identified. Imaging of the plume is carried out by repeating the experiment for different gate delays/time delays ($t_d$) and for a given gatewidth/integration time ($t_w$) using the ICCD. The experiment is also repeated for two different geometries to explore the features of expanding plasma and we compare the same for SP and DP schemes. The DP scheme is implemented using a combination of a polarizing cube beam splitter and a half-wave plate for splitting the energy used in SP scheme into two pulses propagating along different paths. The second pulse travels a longer path, leading to an inter-pulse delay of $\sim$ 800 ps and follows the path of the first pulse after the pulses are combined in a collinear geometry (back to back irradiation at the same position). Also, measurements are repeated for various irradiation energies in order to find out the variation in nature of expansion and changes in the abundance of species.

\section*{Results}
\begin{figure}[ht!]
	\centering
	\includegraphics[width=0.9\linewidth]{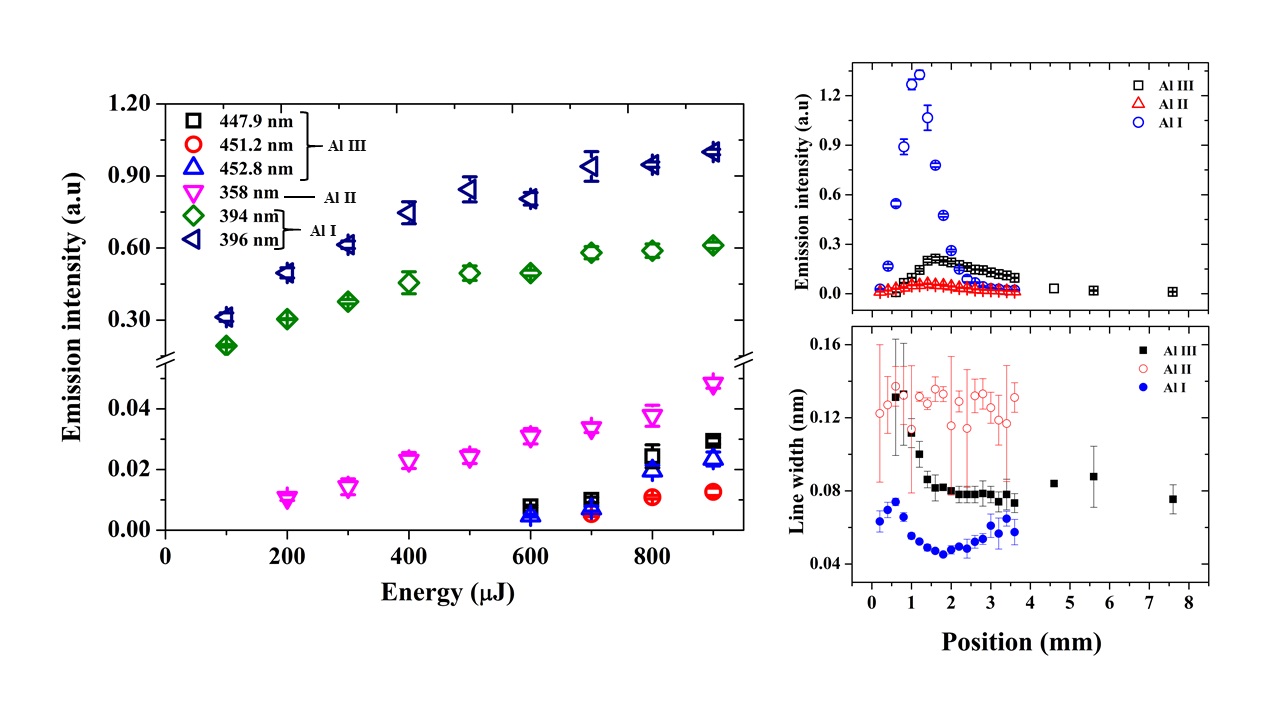}
	\caption{Emission from charged and neutral Al species from the plasma plume measured for various energy of irradiations (left) and for different axial positions for a fixed energy $>$ 600 $\mu$J (right). It is clear from the figure that energies $\geq$ 600 $\mu$J produces Al III species in the plasma meanwhile Al I emission saturates.}
	\label{fig:eipos}
\end{figure}

Time-resolved OES for various energies and for different positions along the expansion direction of the plume is carried out to understand the abundance of species in the plasma. Emission lines at 394 nm and 396 nm (for Al I), 358 nm (Al II) and 447.9 nm, 451.2 nm and 452.8 nm (for Al III) have been used in the present case for analysis. Emission from different species is found to be dependant on the irradiation energies. While only Al I emission is recorded for 100$\mu$J, emission from both Al I and Al II is visible until 500 $\mu$J. Al III emissions are visible along with Al I and Al II for all irradiation energies above 500 $\mu$J which is consistent with the previous reports on the dependence of ion yield on laser fluence\cite{fluence}. It is also found that the emission from Al I saturates once emission from Al III appears in the OES. From OES measurements along the expansion direction for $\sim$ 600 $\mu$J, the emission maximizes at $\sim$ 1 mm above the target surface for all the species. It is found that the emission from Al III has a larger spatial extent compared to Al I and Al II, which can be attributed to the generation of faster species via non-thermal processes (See Fig \ref{fig:eipos}).
%Up to three levels of \textbf{subheading} are permitted. Subheadings should not be numbered.

%\begin{figure}
%	\centering
%	\begin{minipage}{.5\textwidth}
%		\centering
%		\includegraphics[width=1.2\linewidth]{EI.pdf}
%		\end{minipage}%
%	\begin{minipage}{.5\textwidth}
%		\centering
%		\includegraphics[width=.7\linewidth]{POS.png}
%		\end{minipage}
%	\captionof{figure}{Emission from charged and neutral Al species from the plasma plume measured for various energy of irradiations (left) and for different axial positions for a fixed energy $>$ 600 $\mu$J (right). It is clear from the figure that energies $\geq$ 600 $\mu$J produces Al III species in the plasma meanwhile Al I emission saturates.}
%\label{fig:eipos}
%\end{figure}

\begin{figure}[ht!]
	\centering
	\includegraphics[width=0.7\linewidth]{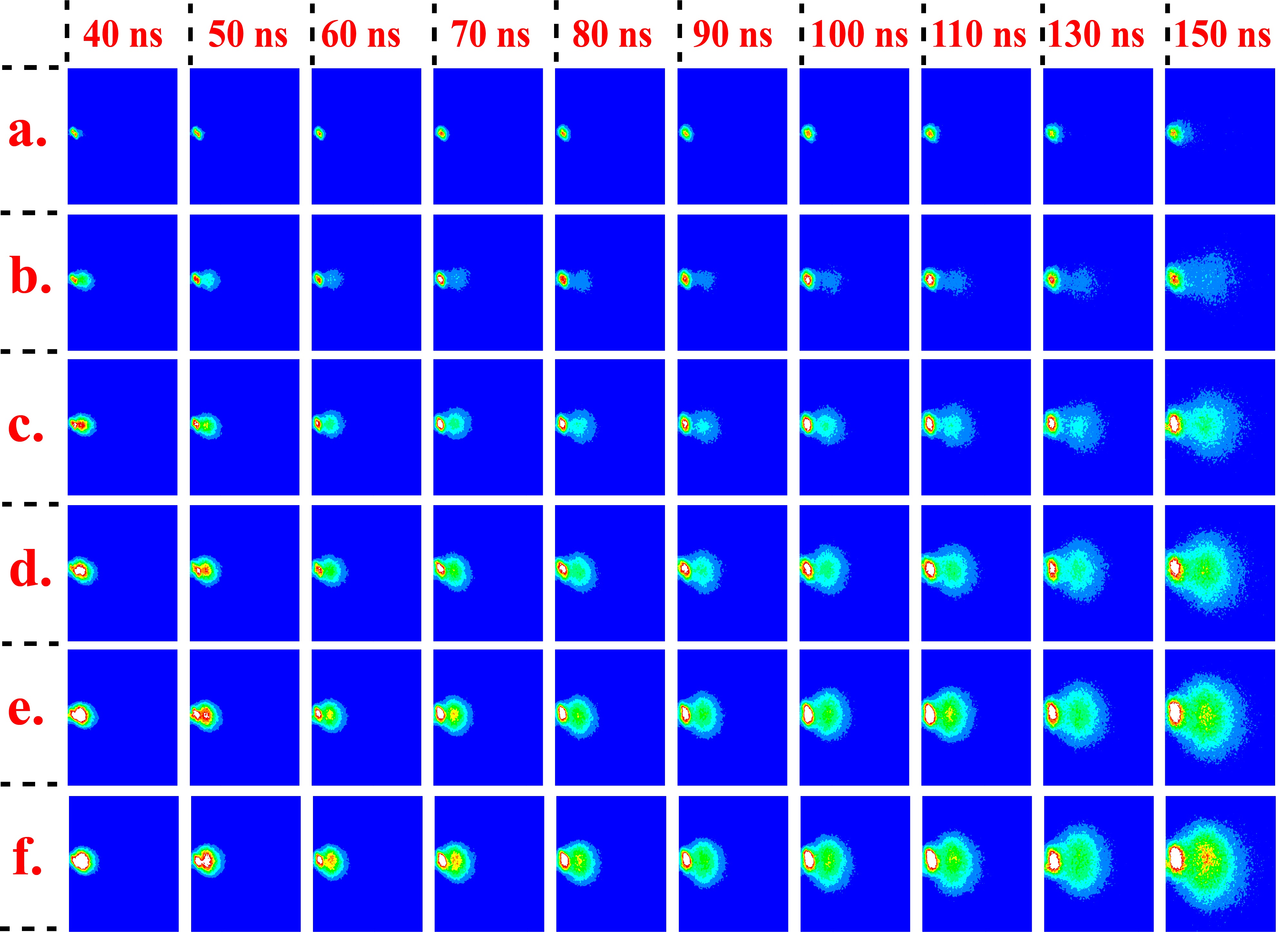}
	\caption{Dynamics of the plasma plume measured for various time delays with 10\% of the time delay as integration time. This is repeated for different irradiation energies of a) 100 $\mu$J, b) 200 $\mu$J, c) 300 $\mu$J d) 400 $\mu$J, e) 500 $\mu$J and f) 600 $\mu$J for single pulse (SP) for various time delays as mentioned in the header of each column. Each image is 20 mm $\times$ 15 mm in dimension.}
	\label{fig:SP}
\end{figure}

\begin{figure}[ht]
	\centering
	\includegraphics[width=0.7\linewidth]{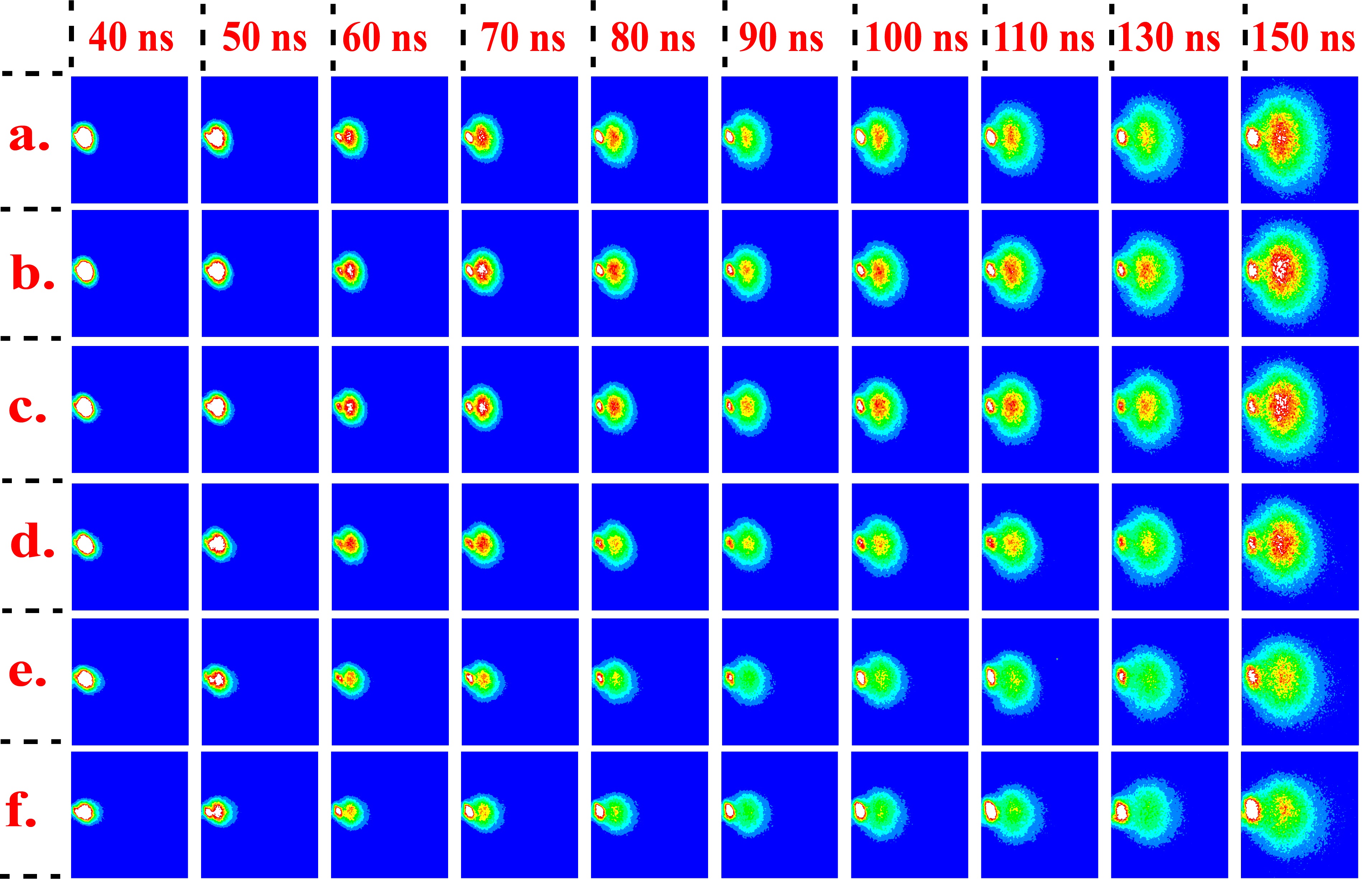}
	\caption{Plume dynamics measured for various time delays with 10\% of the time delay as integration time for different irradiation schemes (dimension of each image is 20 mm $\times$ 15 mm ) like a) 100 $\mu$J-500 $\mu$J, b) 200 $\mu$J-400 $\mu$J and c) 300 $\mu$J-300 $\mu$J, d) 400 $\mu$J -200 $\mu$J, and e) 500 $\mu$J -100 $\mu$J combinations for double pulse (DP) schemes. The image corresponding to SP at 600$\mu$J irradiation which is indicated as \textbf{f} is included for a comparison. }
	\label{fig:DP}
\end{figure}
Measurements on the plume dynamics have been carried out for various $t_d$ s up to 500 ns starting from 40 ns with 10\% of $t_d$ as $t_w$. This has been repeated for energies as shown in Fig. \ref{fig:SP} for a) 100 $\mu$J ,b) 200 $\mu$J, c) 300 $\mu$J, d) 400 $\mu$J, e) 500 $\mu$J and for f) 600 $\mu$J for SP. For 100 $\mu$J, the plume is bright with counts per pixel increasing initially with time, decreasing soon after and eventually diluted to the surrounding  after it has reached a certain distance from the target surface (namely, the plume length) and is $\sim$ 4.3 mm in $\sim$ 110 ns. Also, there is only one peak/component observed for 100 $\mu$J with negligible expansion for the measurement parameters. The hottest spot, the point in the plume at which maximum emission occur, is observed to be at the same spatial point $\sim$ 1.5 mm above the target surface from $t_d$ = 40 ns until the plume gets diluted. On increasing the energy of irradiation to 200 $\mu$J, the plume exhibits two peaks/components, namely the fast and slow peaks. While fast and slow components display equal emission counts for their peaks at earlier times of expansion, their relative emission counts show a significant difference at later times. For all energies $\geq$ 200 $\mu$J,  the emission from slow species is larger for all measured $t_d$s. Moreover, the emission count increases with energy for both fast and slow components, which is evident in the OES measurements (see Fig. \ref{fig:eipos}).  Also, an increase in the plume length has been observed when  plumes corresponding to 100 $\mu$J are compared with 200 $\mu$J which is shown in Fig \ref{fig:SP} \textbf{a} and \textbf{b}. 

The emission features for  DP schemes for various energy combinations such as a) 100 $\mu$J-500 $\mu$J, b) 200 $\mu$J-400 $\mu$J, c) 300 $\mu$J-300 $\mu$J, d) 400 $\mu$J -200 $\mu$J and e) 500 $\mu$J -100 $\mu$J are shown in Fig. \ref{fig:DP}. The inter-pulse delay is fixed to $\sim$ 800 ps since the improvements in ion density has been reported for delays $<$ 1 ns\cite{pinon2009optical}. The plume in DP scheme is found to be entirely different from SP not only in terms of the total emission counts but also the geometrical shape of the expanding plume. In the case of SP, the plasma is more directional with relatively low lateral expansion as indicated in Fig. \ref{fig:SP} showing a trend similar to a femtosecond (fs) plasma \cite{Verhoff2012}. Whereas for DP scheme, the lateral expansion of the plume is evident resembling a ns plasma expansion (as shown in Fig. \ref{fig:DP} \textbf{a}-\textbf{e}) \cite{Verhoff2012}, though the process leading to the shape of the plasma plume may be different. The splitting of the plume into two components; the fast and slow components is evident, but occurs little later, i.e. for larger $t_d$s in all DP cases compared to all the SP schemes. The current observation implies that the fast emitting species are mostly either highly charged species or the result of recombination of the fastest species traveling at very high velocities. This is corroborated with the measured spectra from highly charged species at larger distances (see Fig. \ref{fig:eipos}). Thus, it could be inferred that the fast components consist mostly of species from non-thermal origin whereas the slow component in general exhibits  thermal nature \cite{hassanien2016}.  Further, the emission counts are higher for cases \textbf{a}, \textbf{b} and \textbf{c} of Fig. \ref{fig:DP} compared to all other DP combinations and SP schemes. The variation in geometric shape of the plume for different irradiation schemes is also different. While the shape of the plume changes from cylindrical to more spherical with an increase in irradiation energy for SP, the DP scheme displays a better spherical shape for cases \textbf{a}, \textbf{b} and \textbf{c}  with aspect ratio closer to 1, i.e. $\sim$ 1.4, than all the other cases presented here (please see the Figure 1 in supplementary file for more details). From these observations, it is clear that the plume dynamics shows distinctive features for SP and DP, more precisely, in the geometric shape of the plume and integrated emission intensities from the plasma. The emission intensity profile is also estimated by averaging the signal over several pixel rows from the recorded images to get a complementary dataset that yields information similar to optical time-of-flight (OTOF) measurements except the fact that current measurements provide a convolution of time-of-flights of all species that move within the plume, a representative figure of which is  shown in Fig \ref{fig:TOF} (a more detailed information is available in Figure 2 and Figure 3 of the supplementary file). While the emission intensities for fast species are higher compared to its slow counterpart in all DP cases, the emission intensity of slow components is larger in SP schemes.

\begin{figure}[ht]
	\centering
	\includegraphics[width=0.7\linewidth]{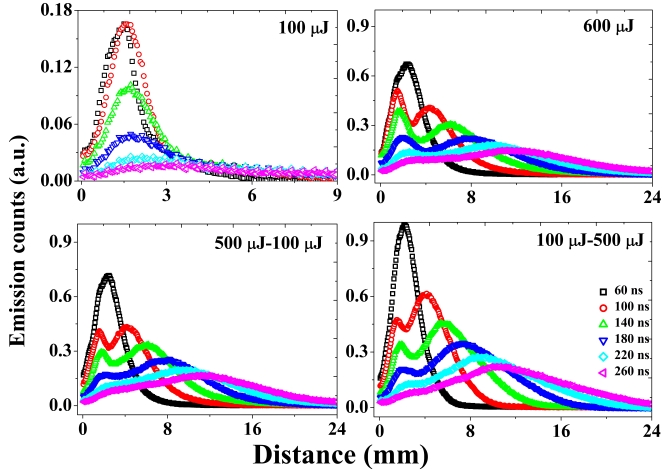}
	\caption{The emission intensity profiles are estimated from averaging the signal over several pixel rows from the recorded images to give a complementary dataset that yields information similar to optical time-of-flight (OTOF) measurements for SP at 100 $\mu$J and 500 $\mu$J and DP at 500 $\mu$J-100 $\mu$J and at 100 $\mu$J-500 $\mu$J. The emission counts are maximum for 100 $\mu$J-500 $\mu$J combination. }
	\label{fig:TOF}
\end{figure}

\begin{table}[ht]
	\centering
	\caption{\bf Velocities of fast and slow components in a single pulse LPP }
	\begin{tabular}{|c|c|c|}
		\hline
			Energy($\mu$J) & Fast (km/s) & slow (km/s)  \\
		\hline
		100 & 				 & 1.41$\pm$0.35 \\
		200 & 44.36$\pm$ 1.28 & 3.19 $\pm$0.55 \\
		300 & 41.92$\pm$1.55  & 5.22$\pm$0.59 \\
		400 & 45.03$\pm$1.37  & 9.42$\pm$0.93 \\
		500 & 43.74$\pm$1.25 & 6.19$\pm$0.65 \\
		600 & 46.29$\pm$0.92  & 10.50$\pm$0.97 \\
		\hline
	\end{tabular}
	\label{tab:table1}
\end{table}

\begin{table}[ht]
	\centering
	\caption{\bf Velocities of fast and slow components in a double pulse LPP }
	\begin{tabular}{|c|c|c|}
		\hline
		Energy($\mu$J-$\mu$J) & Fast (km/s) & slow (km/s)  \\
		\hline
		100-500 & 42.99$\pm$1.37 & 7.07$\pm$0.80 \\
		200-400 & 37.15$\pm$0.69& 7.26$\pm$0.84 \\
		300-300 & 40.79$\pm$1.36  & 5.68$\pm$0.80 \\
		400-200 & 39.74$\pm$1.47 & 4.66$\pm$1.22 \\
		500-100 & 45.85$\pm$1.52 & 6.63$\pm$1.25 \\
		\hline
	\end{tabular}
	\label{tab:table2}
\end{table}

\begin{figure}[ht!]
 	\centering
 	\includegraphics[width=0.8\linewidth]{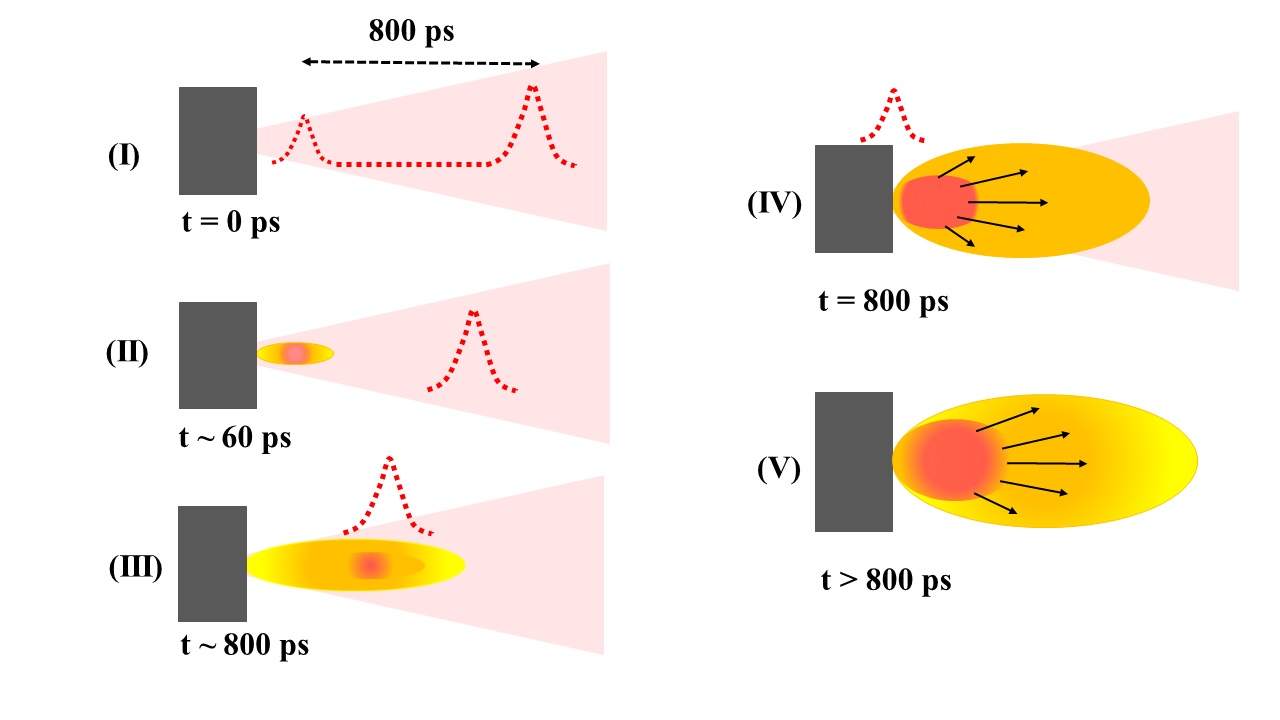}
 	\caption{Schematic for generation of plasma in DP geometry. Here the first pulse produces a plasma (II) and the second pulse delayed by 800 ps then interacts with the plasma produced by the first laser pulse via mechanisms such as (III) laser-plasma, (IV) laser-target and (V) plasma-plasma interactions.}
 	\label{fig:scheme}
 \end{figure}

The velocities of fast and slow components of the plasma calculated for SP and DP cases at later times can be found in Table. \ref{tab:table1} and Table. \ref{tab:table2} respectively. It is clear from Table 1 that the fast components in most cases are moving with nearly the same average velocity irrespective of the irradiation energies except for 100 $\mu$J (where no evidence of fast species is detected). However, velocities of the slow component increase with an increase in energy for SP. Also, from the plume dimensions, plume length is found to be the same for SP at 600 $\mu$J and for all DP cases; although there is a definite change in plume width for each case. To understand this further, the SP at 100 $\mu$J, (Fig. \ref{fig:SP} \textbf{a}) is compared with DP at 100 $\mu$J-500 $\mu$J (Fig.\ref{fig:DP} \textbf{a}.). The plume is found to concentrate along the plume axis at 100 $\mu$J for all $t_w$s and such a plume would expand to $\sim$ 80 $\mu$m from the target surface (given the spot size is $\sim$ 85 $\mu$m) assuming the maximum initial velocity to be $\sim$ 100 km/s in 800 ps. The situation is similar when the first pulse irradiates the target in the DP case \textbf{a}, producing a plasma plume. As soon as the second pulse arrives, the nature of expansion of the plume changes dramatically; the schematic of which is illustrated in Fig. \ref{fig:scheme}. Fig. \ref{fig:scheme} \textbf{(I)} and \textbf{(II)} represents the irradiation of the target before and after the first pulse reaching the surface respectively. A plasma will be formed by the first pulse and the delayed pulse may interact with the plasma via inverse-bremsstrahlung (a process that depends on number density and temperature of the plume). The cross section of this process would be low due to the lower number density in the present case, leading to ineffective shielding of the second pulse from reaching the target surface. This would invoke processes such as: 1) laser-plasma (Fig. \ref{fig:scheme} \textbf{(III)}), 2) laser-target (Fig. \ref{fig:scheme} \textbf{(IV)}) and 3) plasma-plasma interactions (Fig. \ref{fig:scheme} \textbf{(V)}). The second plasma  generated from the heated pit formed by the first pulse would then experience the pressure of a dynamic plasma produced by the first laser pulse. This could be compared to a situation similar to the expansion of a plasma plume at moderate pressures of $\sim$ 1 Torr, except that the pressure exerted by the pre-formed plasma plume is dynamic and can have a multitude of interactions as compared to the static ambient pressure. Hence, the two  plasmas thus produced can interact with each other, producing shock waves and other possible collisional interactions, which make the expansion relatively complex. This results in a plasma plume with larger radial expansion than in SP cases due to enhanced collisions and energy exchange processes. The ablation efficiency from the target surface for cases \textbf{a} to \textbf{c} in DP is larger due to the increased ablation efficiency from a heated molten surface. Therefore it can be concluded that the lateral expansion would be prominent for irradiation schemes in which the first pulse energy is lower/equal to the second pulse. In general, due to the interaction of neutral and charged species in the plasma produced by the first pulse and with fast electrons produced by the second plasma, it is very likely to produce highly charged species in the DP case similar to the situation of larger irradiation energies in SP scheme. 
 
In addition to the above DP scheme, irradiation at two spatial points (DP2) separated by 700 $\mu$m on the target was also performed and showed distinctive features compared to SP and the above mentioned collinear DP cases (which will be called DP1 hereafter). Two pulses of equal energies (400 $\mu$J/pulse) with zero delay between pules are used for DP2. Though an adiabatic expansion of plasma with the existence of two peaks (fast and slow) is evident in all cases, the plume expansion is found to have considerable variation in its shape for DP2 due to the interaction of the two plasmas, formed upon the irradiation at two different spatial points on the target surface, forming colliding plasmas.  The plume expansion is found to be confined along  the plume axis leading to a more cylindrical plasma plume that resembles the expansion features of a fs plasma plume (please see Figure 4 in the supplementary file for more details). The slow peaks are broadened spatio-temporally leaving the possibility for the generation of nanoparticles at a later stage. A more detailed study would be necessary in this regard as this method can be used to create more uniform thin films as the plume is more cylindrical compared to fs plasma.

%\subsection*{Subsection}
%Example text under a subsection. Bulleted lists may be used where appropriate, e.g.
%\begin{itemize}
%\item First item
%\item Second item
%\end{itemize}
%\subsubsection*{Third-level section}
 %Topical subheadings are allowed.
\section*{Discussion}

A picosecond laser plasma has been generated and characterized to investigate the dynamics of the plume so as to make it suitable for various applications in which the plume shape as well as abundance of certain species are crucial. From these measurements it is demonstrated that the features of the plume  are governed and modified by the double pulse geometry, such that homogeneity of the plume could be enhanced, i.e. for $t_d$ $\leq$ 60 ns from 100 $\mu$J-500 $\mu$J to 300 $\mu$J-300 $\mu$J, at least in the earlier stages of expansion before plume splitting due to fast and slow species occurs. A detailed analysis of laser-plasma, laser-target and plasma-plasma interaction for the double pulse geometry is performed to understand the structural modifications in the plume. Optical emission spectroscopy facilitated the understanding of the abundance of species both spatially and temporally for given energies, while imaging with an ICCD was used to investigate the hydrodynamics of the plasma plume. From these measurements it is clear that the structure of the plasma plume is modified by employing a double pulse geometry which has been analyzed carefully to understand the lateral expansion of the plume. While the scheme DP2 is predicted to favor generation of nanoparticles at a later stage, the plume in DP1 scheme exhibits symmetry in expansion showing a homogeneity useful for applications such as high-order harmonic generation that demand phase-matching.

\section*{Methods}

A plasma is generated using 60 ps laser pulses at 800 nm from a mode-locked Ti:Sapphire  laser (\textit{Odin II}, Quantronix) focused onto a solid Al target, which is kept in nitrogen ambient at pressure $\sim$ 10 $^{-6}$ Torr. Optical emission spectroscopy and plasma plume dynamics have been recorded using a high resolution spectrometer equipped with Gen II ICCD (\textit{Pi: MAX 1024 f}, Princeton Instruments). Imaging of the plume is carried out by repeating the experiment for different gate delays/time delays and for a given gate width/integration time. The experiment is also repeated for two different geometries, single pulse and double pulse, to explore and compare the features of an expanding plasma plume.

\bibliography{sample}

%\noindent LaTeX formats citations and references automatically using the bibliography records in your .bib file, which you can edit via the project menu. Use the cite command for an inline citation.

\section*{Acknowledgements}

This project is funded by the Australian Research Council Linkage project grant No. LP140100813. N. Smijesh has been supported by the Griffith University Postdoctoral Fellowship Scheme. Kavya H. Rao and D. Chetty were supported through an ``Australian Government Research Training Program Scholarship".

\section*{Author contributions statement}

N. Smijesh, I.V. Litvinyuk and R.T. Sang conceived the experiments, supervised data analysis. N. Smijesh, Kavya H. Rao and D. Chetty conducted the experiments, collected and anlalyzed data.. R.T Sang and I. V. Litvinyuk directed the research, provided the resorces. All authors were involved in discussion and reviewing of the manuscript. 

\section*{Additional information}

The author(s) declare no competing financial interests.

%\begin{figure}[ht]
%\centering
%\includegraphics[width=\linewidth]{stream}
%\caption{Legend (350 words max). Example legend text.}
%\label{fig:stream}
%\end{figure}

%\begin{table}[ht]
%\centering
%\begin{tabular}{|l|l|l|}
%\hline
%Condition & n & p \\
%\hline
%A & 5 & 0.1 \\
%\hline
%B & 10 & 0.01 \\
%\hline
%\end{tabular}
%\caption{\label{tab:example}Legend (350 words max). Example legend text.}
%\end{table}

%Figures and tables can be referenced in LaTeX using the ref command, e.g. Figure \ref{fig:stream} and Table \ref{tab:example}.

\end{document}